\documentstyle[12pt,epsfig]{article}

\makeatletter
\def\section{\@startsection {section}{1}{\z@}{3.5ex plus 1ex minus
    .2ex}{2.3ex plus .2ex}{\sc }}
\def\subsection{\@startsection{subsection}{2}{\z@}{3.25ex plus 1ex minus
   .2ex}{1.5ex plus .2ex}{\small \sc }}
\def\subsubsection{\@startsection{subsubsection}{2}{\z@}{3.25ex plus 1ex minus
   .2ex}{1.5ex plus .2ex}{\small \sc }}
\def\appendix{\par\clearpage
  \setcounter{section}{0}
  \setcounter{subsection}{0}
  \@addtoreset{equation}{section}
  \def\@sectname{Appendix~}
  \def\theequation{\thesection.\arabic{equation}}
  \def\thesection{\Alph{section}}}
\makeatother

\newcommand{\pss}{\protect\scriptscriptstyle}

\begin{document}

\begin{flushright}
BUTP-97/12\\
Bern, July 1997
\end{flushright}

\vskip 2. truecm

\begin{center}{\Large\bf Some $\textstyle\protect\rm\bf O(\alpha^2)$ 
Annihilation Type Contributions\\
to the Orthopositronium Width.}
\end{center}

\vskip 1. truecm
\centerline{V. Antonelli$^a$, V. Ivanchenko$^b$, E. Kuraev$^c$ and V. 
Laliena$^a$.}

\vskip 1 truecm
\centerline {\it $^a$ Institute for Theoretical Physics, University of Bern.}
\centerline {\it Sidlerstrasse 5, 3012 Bern (Switzerland).}
\vskip 0.15 truecm
\centerline {\it $^b$ Institute of Nuclear Physics, Siberian Division}
\centerline{630090 Novosibirsk, Russia}
\vskip 0.15 truecm
\centerline {\it $^c$ JINR, 141980 Dubna, Russia}
\vskip 0.3 truecm
\centerline{e-mail : antonell@butp.unibe.ch  \  V.N.Ivanchenko@inp.nsk.su} 
 
\hspace*{2.1 truecm}
kuraev@thsun1.jinr.dubna.su \ laliena@butp.unibe.ch 
\vskip 2 truecm

\centerline{ ABSTRACT}

\vskip 0.5truecm

\noindent
We consider some radiative corrections to the lowest order annihilation 
diagram for the orthopositronium decay rate. 
The insertion of the renormalized vertex correction in 
the annihilation graph gives $1.6283 \ (\alpha/\pi)^2\,
\Gamma_0$. We compute also the contribution of the square of the lowest order 
annihilation  amplitude, which turns out to be 
$0.1702\,(\alpha/\pi)^2\,\Gamma_0$.
Finally, we obtain a term $\alpha^2\,\ln\alpha\,\Gamma_0$ arising from the 
correction to the light--light scattering block due to the exchange of one 
coulombic photon, in agreement with earlier computations.

%S**********
%\end{titlepage}
%S***********

\newpage

\section{\bf Introduction}

Measurements of orthopositronium (${\textstyle\protect\rm O_{ps}}$) 
decay rate in the recent 
years pose a great challenge in QED due to the large discrepancy between
the experimental result and the theoretical predictions \cite{review}. 
In fact, the two most accurate experimental rates \cite{exp1,exp2}

\begin{displaymath}
\lambda_{O_{ps}}^{exp}\;=\;7.0514\,\pm\,0.0014\;\mu s^{-1}\;\;\;\;\;\;
{\textstyle\protect\rm and}\;\;\;\;\;\;
\lambda_{O_{ps}}^{exp}\;=\;7.0482\,\pm\,0.0016\;\mu s^{-1}\, ,
\end{displaymath}

\noindent
deviate by $9.4\,\sigma$ and $6.2\,\sigma$ from the theoretical
value\footnote{Recently a new experiment \cite{newexp} gave the value
$7.0398\pm0.0025\pm0.0015 \,\mu s^{-1}$ (the first error is statistical
and the second systematic), in good agreement with the theoretical 
expectation. However, an independent confirmation of this measurement
is necessary before concluding that the ${\textstyle\protect\rm O_{ps}}$ 
problem is experimental instead of theoretical. \label{note1}}, 
whose most accurate estimate to order $\alpha / \pi$ has been given in 
\cite{adkins1}: 

\begin{displaymath}
\lambda_{O_{ps}}^{\protect\scriptscriptstyle QED} (PS\rightarrow 3\,\gamma)
\;=\;7.038236\,\pm\,0.000010\,\mu s^{-1} \ .
\end{displaymath}

\noindent
The decay rate to leading order,

\begin{equation}
\Gamma_0\;=\;\frac{\alpha^6 m c^2}{\hbar}\,\frac{2\,(\pi^2 - 9)}{9\,\pi}
\;=\;7.21117\,\mu s^{-1}\, ,
\end{equation}

\noindent
 was computed by Ore and Powell 
\cite{op} and the order $\alpha$ corrections by several authors 
\cite{pascual, stroc, lepage, adkinsap, adkins1}.

Faced up to this difficulty, theorists have made a great 
 effort to compute the next correction, $\left(\alpha/\pi \right)^2$, of the
perturbative expansion \cite{adkins3, adkins4, burichenko}:

\begin{eqnarray}
\lambda_{O_{ps}}^{\scriptscriptstyle QED}\,(PS\rightarrow 3\,\gamma) &=&
\Gamma_0\,
\left[\,1\:+\:(-10.2866\pm0.0006)\,\frac{\alpha}{\pi}\right. \nonumber \\
&+&\left. C\,\left(\frac{\alpha}{\pi}
\right)^2\:+O((\alpha /\pi)^3)\,\right]\, .
\end{eqnarray}

\noindent
To get rid of the theoretical--experimental discrepancy, the coefficient 
$C$ must be of order
$250\pm40$. Such a large value cannot be excluded, even if it
may appear unnatural in the framework of perturbation theory (PT).
If the results of \cite{newexp} are correct (see footnote \ref{note1})
only $C\approx 30$ is required. 

The computation of $C$ is very hard due to the large
number of Feynman diagrams contributing to the $\alpha^2$ order of PT.
Some of these have been already calculated:
the vacuum polarization type corrections to the
first order graphs were considered in
\cite{adkins3}, the radiative corrections to the light--light scattering
block in \cite{adkins4}, and the square of the first order
amplitude  in \cite{burichenko}. The relativistic
corrections, i.e. those associated with the expansion in $v/c\sim\alpha$, 
where $v$ is the relative velocity of the $e^+e^-$ pair in positronium,
were taken into account up to order $\alpha^2$ in ref. \cite{relat}.

All the contributions to the amplitude up to second order PT may be written
in the form:

\begin{eqnarray}
M &=& M_0\;+\;\frac{\alpha}{\pi}\,\left(\,M_{\pss B'}\:+\:M_{\pss A}\:+\:
M_1\,\right)\;+ \nonumber \\
& & \left(\frac{\alpha}{\pi}\right)^2\,\left(\,M_{\pss AB'}\:+\:M_{\pss AR}
\:+\:M_2\,\right)\;+\;0\left(\alpha^3\right)\,,
\end{eqnarray}

\noindent
where\footnote{Note that here and in the rest of the paper we write 
explicitly the powers of $\alpha /\pi$, relative to the lowest
order $M_0$, for each amplitude. 
(For example, $M_{\pss A}$ is of the same order as $M_0$). On the contrary, we
omit them in the text. Note also that we 
will not write any power of $\alpha /\pi$ for the two unsubtracted binding
amplitudes, 
$M_{\pss B}$ and $M_{\pss AB}$, since they contain terms of different 
order in $\alpha /\pi$.} 
$M_1$ represents the sum of all the
first order amplitudes with the exceptions of the annihilation diagram (see 
fig. 1), denoted by $M_{\pss A}$, and the subtracted binding amplitude, 
$M_{\pss B'}$ \cite{adkins1}. The second
order annihilation type corrections  are given by the 
subtracted binding diagram, $M_{\pss AB'}$ (fig. 2--A) and the radiative 
corrections to the light--light scattering block, $M_{\pss AR}$ (an example of
which is given in fig. 2--B). 
$M_2$ denotes the remaining (non--annihilation type) second order 
amplitudes.

In this paper we consider the second order corrections given by $M_{\pss A}$
and $M_{\pss AB'}$ and the logarithmic enhancement produced by a
coulombic one photon exchange in $M_{\pss AR}$.
The contribution of these corrections to the decay rate has the form  

\begin{equation}
\left(\frac{\alpha}{\pi}\right)^2\,\left[\,2\,{\textstyle Re}\,
(\,M_{\pss AB'}^*\,M_0\,)\:+\:\left|M_{\pss A}\right|^2\,\right]\:+\:
\alpha^2\,\ln\alpha\,\left|M_0\right|^2\, .
\end{equation}

\noindent
The previous expression must be summed over the final photon polarizations,
averaged over the $\protect\textstyle\rm O_{ps}$ spin states and
integrated over the phase space of the three final photons, with the
proper kinematical factors (see for example \cite{adkins1}).
 
The remaining part of the paper is organized as follows: in the second section 
we compute the contributions of the binding corrections to the 
lowest order annihilation diagram. 
In section three, using the known results for the
light--light scattering tensor, we compute the contribution of the 
square of the lowest order annihilation amplitude. Finally, in the fourth
section we consider the exchange of one coulombic photon in the light--light
scattering block of the annihilation amplitude, finding a logarithmic
contribution in agreement with earlier calculations 
\cite{lepage, logrus, adkins4}.

\section{\bf Radiative Corrections to the Annihilation Process}

It is well known \cite{adkins1, tomo} that the contribution to
the amplitude for the $\textstyle\protect\rm O_{ps}$ decay rate
originated by the binding diagram contains also the lowest order 
approximation:

\begin{equation}
M_{\pss B}\;=\;M_0\:+\: \frac{\alpha}{\pi} M_{\pss B'}\, .
\end{equation}

\noindent
Therefore only the subtracted binding diagram $M_{\pss B'}$ must be included, 
otherwise $M_0$ would be counted twice. This phenomenon
occurs due to the presence of the coulombic part in the
virtual photon propagator, which had been already taken into account when 
solving the Bethe--Salpeter equation for the $\textstyle\protect\rm O_{ps}$ 
wave function.

It is quite clear that the analogous phenomenon should come out in the 
``binding'' type radiative correction to the lowest order annihilation 
diagram (see fig. 2--A), and in fact  we shall show
in this section that the amplitude $M_{\pss A}$ is contained in $M_{\pss AB}$. 
Therefore, we can write:

\begin{equation}
M_{\pss AB}\;=\; \frac{\alpha}{\pi}\, M_{\pss A}\:+\: 
\left(\frac{\alpha}{\pi}\right)^2 \,M_{\pss AB'}\, ,
\end{equation}

\noindent
where $M_{\pss AB'}$ is the subtracted binding--annihilation amplitude.

We express the amplitude as a product

\begin{equation}
M_{\pss AB}^{\pss (m,\,\lambda)}\;
=\;\frac{-i}{4m^2} \,T^{{\pss (m)}\,\rho}\,G_\rho^{\pss (\lambda)}\ .
\label{mab}
\end{equation}

\noindent
In the previous formula the 4--vector $G_{\rho}^{\pss (\lambda)}$ describes the
transition of the heavy photon to three real ones,  
$\lambda=(\lambda_2,\lambda_3,\lambda_4)$ stands for the
set of the three polarizations of the final photons, $\lambda_i=\pm 1$, and 
$T_{\rho}^{\pss (m)}$ is the order $\alpha$ correction to the annihilation 
current 4--vector of the positronium in the polarization state 
$\vec\epsilon_m$. Explicitly:

\begin{displaymath}
 T_{\rho}^{\pss (m)} =  -\frac{ie\alpha}{4\pi}\int\frac{d^4p}{(2\pi)^4}
\int\frac{d^4k}{i\pi^2}\,\frac{\Delta_{\mu\nu}(k-p)}{(k-p)^2} \:\times
\end{displaymath}
\begin{equation}
\frac{Tr\,\left\{\,\Psi^{\pss (m)}(p)\,
\gamma_{\mu}\,\left[\,-\hat{P}/2+\hat{k}+m\,\right]\,\gamma_{\rho}\,
\left[\,\hat{P}/2+\hat{k}+m\,\right]\,\gamma_{\nu}\,\right\}}
{\left[\,(-P/2+k)^2-m^2\,\right]\left[\,(P/2+k)^2-m^2\right] }
\;\:+\;\:C_t\, , \label{binan}
\end{equation}

\noindent
where $C_t$ stands for the contribution of the vertex counterterm.
We use the notation $\hat{k}=k_\mu\gamma^\mu$ and here and everywhere in
this paper a term $i0$ must be 
implicitly understood in each factor of the denominator arising from a
propagator (only in eq. (\ref{i0eq}) the $i0$ will be explicitly written). 
In eq. (\ref{binan}) $P$ is the $\textstyle\protect\rm O_{ps}$ 
momentum in its rest frame:
\vskip 0.25 cm
\centerline{
$P=(2W,0,0,0)\, , \;\;\;\;{\textstyle\protect\rm with}\;\;\;\; 
W\approx m-\frac{\gamma^2}{2m}\,,\;\;\;\;{\textstyle\protect\rm and} \;\;\;\; 
\gamma = \frac{m\,\alpha}{2}\, $.}
\vskip 0.25 cm
\noindent
The wave function $\Psi^{\pss (m)}(p)$ of $\textstyle\protect\rm O_{ps}$ 
relevant to our approximation is (see for example \cite{adkins1, adkins5}): 

\begin{equation}
\Psi^{\pss (m)}(p) \;=\;(2\pi)\,\delta(p_0)\,\sqrt{2m}\:
\left[ \begin{array}{cc} 
0 & \vec{\sigma}\cdot\vec{\epsilon_m} \\
0 & 0
\end{array} \right]\:\Phi(\vec{p})\, , \label{pwf}
\end{equation}

\noindent
with $\Phi(\vec{p})$ the nonrelativistic ground--state wave 
function 

\begin{equation}
\Phi(\vec{p})\;=\;\phi_0\,\frac{8\pi\gamma}
{\left(\,|\vec{p}|^2
\,+\,\gamma^2\,\right)^2}
\end{equation}

\noindent
and the constant $\phi_0$ is the wave function at the origin,
$\phi_0=\sqrt{\gamma^3/\pi}$.

The $\Delta_{\mu\nu}$ tensor obviously depends on the gauge we use. The choice
of the 
gauge is subtle when dealing with  bound state problems. It has been discussed
elsewhere (see for example \cite{adkinsap}) that the Coulomb gauge is the most
natural for calculations in positronium. However, covariant gauges are
simpler for 
computing radiative corrections, and, among them, the Fried--Yennie (FY) gauge
is the most convenient, due to its good infrared behaviour. We shall compute
$T^{\pss (m)}_\rho$
both in the FY gauge and in the Coulomb gauge. As expected, the result is the
same in both cases, and no gauge correction term must be added when using the
FY gauge.

\subsection{Fried--Yennie gauge.}

The FY gauge is a covariant gauge defined by

\begin{equation}
\Delta_{\mu\nu}(k)\;=\;g_{\mu\nu}\:+\:2\,\frac{k_\mu k_\nu}{k^2}\, .
\label{fygauge}
\end{equation}

\noindent
It has good infrared properties, allowing us to work safely with a
zero fictitious photon mass from the beginning. 

Following \cite{adkins1} we separate the trace entering the integrand of
(\ref{binan}) into two pieces, 
one which remains non--singular at k=0 and one containing the contribution
of the coulombic photon. To this end, we define

\begin{equation}
tr_{\mu\nu\rho}(k)\;=\;Tr\,\left\{\,\Psi^{\pss (m)}(p)\,
\gamma_{\mu}\,\left[\,-\hat{P}/2+\hat{k}+m\,\right]\,\gamma_{\rho}\,
\left[\,\hat{P}/2+\hat{k}+m\,\right]\,\gamma_{\nu}\,\right\}\, 
\end{equation}

\noindent
and write: $tr_{\mu\nu\rho}(k) = tr_{\mu\nu\rho}(0) + \left\{tr_{\mu\nu\rho}(k)
- tr_{\mu\nu\rho}(0)\right\}$. 
\vskip 3mm

We consider first the term $tr_{\mu\nu\rho}(0)$ . The $\gamma$ matrices algebra
leads to:

\begin{equation}
tr_{\mu\nu\rho}(0)\;=\;-4\,W^2\,\delta_{\mu 0}\,\delta_{\nu 0}\,Tr\left\{
\gamma_\rho\,
\left[ \begin{array}{cc} 
0 & \vec{\sigma}\cdot\vec{\epsilon_m} \\
0 & 0
\end{array} \right]\,\right\}\;+\;O(\alpha^2)\, .
\end{equation}

\noindent
To arrive to the last expression we have used the fact that

\begin{equation}
\left(\frac{1}{2}\hat{P}\,-\,m\right)\,\Psi^{\pss (m)}(p)\;=\; O(\alpha^2)
\;\;\;\;\;\;\;\;
\Psi^{\pss (m)}(p)\,\left(\frac{1}{2}\hat{P}\,+\,m\,\right) \;=\; O(\alpha^2)\, 
.
\label{positde}
\end{equation}

The integration in $p_0$ in (\ref{binan}) is trivial,
using the delta function entering the formula (\ref{pwf}). It remains the 
following integral:

\begin{equation}
\int\,\frac{d^3p}{(2\pi)^3}\,\frac{8\pi\gamma}
{\left(|\vec{p}|^2+\gamma^2\right)^2}\,\int\,\frac{d^4 k}{(i\pi^2)}\,
\frac{-W^2\,\left(1+2\,\frac{k_0^2}{(k-p)^2}\right)}
{(k-p)^2\left[(k+\frac{1}{2}P)^2-m^2\right]
\left[(k-\frac{1}{2}P)^2-m^2\right]}\ ,
\label{adkinsint}
\end{equation}

\noindent
whose result, $\frac{\pi}{\alpha} - 3\:+\:O(\alpha^2)$, can be found in 
\cite{adkins1}. Hence the contribution of this term to $T^{\pss (m)}_\rho$
is:

\begin{equation}
- \, i\,\frac{\alpha^3m^2}{\pi}\,Tr\left\{\,\gamma_\rho\,
\left[ \begin{array}{cc} 
0 & \vec{\sigma}\cdot\vec{\epsilon_m} \\
0 & 0
\end{array} \right]\,\right\}\,\left(\,\frac{\pi}{\alpha}\,-\,3\,\right)\ .
\label{conbin}
\end{equation}

Let us now consider the remaining term, 
$tr_{\mu\nu\rho}(k) - tr_{\mu\nu\rho}(0)$.
In this case the integral is free from infrared singularities and we can
put $p=0$ in the loop integral, introducing an error of order $\alpha^2$. 
However, ultraviolet divergences are present;  we regulate them by using 
dimensional regularization (the analogous calculation in 
cut-off regularization is performed in appendix A).
It is important to remember that some care is necessary when using dimensional
regularization in the FY gauge.
As was shown by G. Adkins \cite{adkins2}, it is convenient to choose the tensor
$\Delta_{\mu\nu}$ as:

\begin{equation}
\Delta_{\mu\nu}(k)\;=\;g_{\mu\nu}\:+\:\frac{2}{1-2\epsilon}\,
\frac{k_\mu\, k_\nu}{k^2}\, ,
\end{equation}

\noindent
where $\epsilon=(4-d)/2$ and $d$ is the complex space-time dimension.

Using $\gamma$ matrices algebra and (\ref{positde}) one can see that the only
contribution of  $tr_{\mu\nu\rho}(k) - tr_{\mu\nu\rho}(0)$ to the integral
(\ref{binan}) is given by:

\begin{equation}
Tr \left\{\,\gamma_\mu\,\hat{k}\,\gamma_\rho\,\hat{k}\,\gamma_\nu
\left[ \begin{array}{cc} 
0 & \vec{\sigma}\cdot\vec{\epsilon_m} \\
0 & 0
\end{array} \right]\,\right\}\, .
\end{equation}
  
\noindent
In this way we arrive to the following integral:

\begin{equation}
-\,\frac{i\,e\,\alpha}{4 \pi}\,\int\,\frac{d^d k}
{i\pi^2(2\pi\mu)^{-2\epsilon}}\,\frac{\gamma_\mu\,\hat{k}\,
\gamma_\rho\,\hat{k}\,\gamma_\nu\, \Delta_{\mu\nu}(k)}
{k^2\,\left[(k+\frac{1}{2}P)^2-m^2\right]\left[(k-\frac{1}{2}P)^2-m^2
\right]}\, , \label{kk}
\end{equation}

\noindent
where $\mu$ is the dimensional parameter introduced in dimensional
regularization. The integral (\ref{kk}) can be evaluated by standard 
techniques, giving

\begin{equation}
-\frac{i e \alpha}{4\pi}\,\left(\,3\,D\:+\:8 - 2\,\delta_{\rho 0}\right)\,
\gamma_\rho\, , \label{remaing}
\end{equation}

\noindent
where $D=\frac{1}{\epsilon}-\gamma_E+\ln\frac{4\pi\mu^2}{m^2}$. (The number
$\gamma_E=0.57721$ is the Euler constant).

Note that current conservation implies
$P^\rho\,G_\rho^{\pss (\lambda)}\, = \,0$. Since we work in the positronium 
rest frame ($\vec{P}=0$), it follows $G_0^{\pss (\lambda)}=0$
and the term with $\delta_{\rho 0}$ in (\ref{remaing}) can be ignored.
The contribution of the remaining term to $T^{\pss (m)}_\rho$ is then:

\begin{equation}
-\frac{i \alpha^3}{4\pi}\, m^2 \left(\,3\,D\:+\:8\,\right)\,
Tr\,\left\{\,\gamma_\rho\,\left[ \begin{array}{cc} 
0 & \vec{\sigma}\cdot\vec{\epsilon_m} \\
0 & 0
\end{array} \right]\,\right\} \, . \label{conrem}
\end{equation}

Adding (\ref{conbin}) and (\ref{conrem}), inserting the result in (\ref{mab})
 and noting that the first order
annihilation amplitude $M_{\pss A}$ can be written as

\begin{equation}
\frac{\alpha}{\pi} M_{\pss A}\;=\;-\frac{\alpha^2}{4}\,Tr\,\left\{\,
\gamma_\rho\,\left[ \begin{array}{cc} 
0 & \vec{\sigma}\cdot\vec{\epsilon_m} \\
0 & 0
\end{array} \right]\,\right\}\,G^{{\pss (\lambda)}\,\rho}
(k_2,k_3,k_4) \, , 
\label{cicci}
\end{equation}

\noindent
we obtain the following contribution to $M_{\pss AB}\,$:

\begin{equation}
\frac{\alpha}{\pi}\, M_{\pss A}\, 
\left[1\:+\,\frac{\alpha}{\pi}\: \left(-3\,+\:\frac{3D+8}{4}\,\right)\right]\, . 
\label{boad}
\end{equation}

The contribution of the vertex counterterm will cancel the divergence 
appearing in (\ref{boad}). Due to the Ward identity, 
this counterterm can be obtained from the
self--energy correction to the electron propagator. 

To order $\alpha$ the mass operator has the form:

\begin{equation}
\Sigma(l)\;=\;\frac{\alpha}{4\pi}\,\int\frac{d^4k}{i\pi^2}\,
\frac{\gamma_{\mu}\,(\hat{l}-\hat{k}+m)\,\gamma_{\nu}\,\Delta_{\mu\nu}(k)}
{k^2\,\left[(l-k)^2-m^2\right]}\, . \label{se}
\end{equation}

\noindent
On mass--shell renormalization conditions imply

\begin{equation}
\frac{i}{\hat{p}-m_0-\Sigma(p)}\;\rightarrow\;\frac{iZ}{\hat{p}-m}\ ,
\;\;\;\hat{p}\rightarrow m \ ,
\end{equation}

\noindent
where $m_0$ is the bare electron mass. From (\ref{se})
it is possible to obtain the electron wave function renormalization constant,
which in the FY gauge is \cite{adkins2}:

\begin{equation}
Z_{FY} \;=\; 1\;-\;\frac{\alpha}{4 \pi}\,\left(\,3\,D\:+\:4\,\right)\ .
\end{equation}

\noindent
It follows that the contribution to the amplitude of the vertex 
counterterm is:   

\begin{equation}
-\left(\frac{\alpha}{\pi}\right)^2\,\frac{3D+4}{4}\,M_{\pss A}\, .
\end{equation}

Finally, summing up all the contributions considered here
(the terms $tr_{\mu\nu\rho}(0)$ and $tr_{\mu\nu\rho}(k) -  tr_{\mu\nu\rho}(0)$,
and the counterterm insertion) we get:

\begin{equation}
M_{\pss AB}\;=\;
\left(\,1\:-\:2\,\frac{\alpha}{\pi}\,\right)\,\frac{\alpha}{\pi}
M_{\pss A}\, . \label{dres}
\end{equation}

\subsection{Coulomb gauge.}

We will show now that the same result is obtained working in the Coulomb
gauge. This gauge is obtained by making the substitution:

\begin{equation}
\frac{-i}{(k-p)^2}\,\Delta_{\mu\nu}(k-p)\;\rightarrow\;
G_{\mu\nu}(k-p) \, ,
\end{equation}

\noindent
where $G_{\mu\nu}(q)$ is the Coulomb propagator:

\begin{equation}
G_{00}(q)\,=\,\frac{i}{\vec{q}^{\, \, 2}}\, ,\;\;\;\;
G_{0i}(q)\,=\,G_{i0}(q)\,=\,0\, ,\;\;\;\;
G_{ij}(q)\,=\,\frac{i}{q^2}\,\left(\delta_{ij}\,-\,
\frac{q_i q_j}{\vec{q}^{\, \, 2}}\right)\, .
\end{equation}

Again, using $\gamma$ matrices algebra and (\ref{positde}), we rewrite the
trace in the numerator of (\ref{binan}) as

\begin{displaymath}
tr_{\mu\nu\rho}(k)\;=\;-P_\mu P_\nu\,Tr\,\left[\,\Psi^{\pss (m)}(p)\,
\gamma_\rho\,\right]\:+\:Tr\,\left[\,\Psi^{\pss (m)}(p)\,\gamma_\nu\,\hat{k}
\,\gamma_\rho\, \hat{k} \, \gamma_\mu\,\right]
\end{displaymath}
\begin{equation}
+\:Tr\,\left[\,\Psi^{\pss (m)}(p)\,\left(\,P_\mu\,\gamma_\nu\,\hat{k}\,
\gamma_\rho\:
-\:P_\nu\,\gamma_\rho\,\hat{k}\,\gamma_\mu\,\right)\,\right]\, .
\label{totalcoul}
\end{equation}

\noindent
It is easy to see that the last term vanishes after contracting the Lorentz 
indices with the ones of the photon propagator. The contribution of the first
term to $T^{\pss (m)}_\rho$ is

\begin{equation}
-\alpha^3\,(16\pi)\,m^4\,Tr\,\left\{\,\gamma_\rho\,\left[ \begin{array}{cc} 
0 & \vec{\sigma}\cdot\vec{\epsilon_m} \\
0 & 0
\end{array} \right]\,\right\}\,I_0\, ,
\end{equation}

\noindent
where $I_0$ is defined by:

\begin{eqnarray}
I_0\;& = & \;\int\,\frac{d^3p}{(2\pi)^3}\,
\frac{8\pi\gamma}{\left(\vec{p}^2\,+\,\gamma^2\right)^2}\, \times \nonumber\\
 & &\int\,\frac{d^4 k}{(4\pi)^2}\,\frac{1}{\left(\vec{k}-\vec{p}\right)^2\,
\left[(-P/2+k)^2-m^2\right]\,\left[(P/2+k)^2-m^2\right]}\, .
\end{eqnarray}

\noindent
This integral has been studied in \cite{tomo}. Its result is

\begin{equation}
I_0\;=\;\frac{i}{(4\pi)^2}\,\frac{1}{m^2}\,
\left(\,\frac{\pi}{\alpha}\:-\:2\,\right) \, .
\end{equation}

\noindent
Using this result and considering (\ref{mab}) and (\ref{cicci}), we find that 
the contribution of the first term of (\ref{totalcoul}) to the
amplitude is 

\begin{equation}
(\,1\,-\, 2 \, \alpha/\pi\,)\, \left(\frac{\alpha}{\pi} \ M_A\, \right) , 
\end{equation}

\noindent
which is the total result obtained in the FY gauge.

Now we will show that the contribution of the second term in (\ref{totalcoul})
exactly cancels against the contribution of the Coulomb gauge vertex
counterterm. Hence, the result in the Coulomb gauge is the same as in the FY
gauge.

The second term of (\ref{totalcoul}) gives raise to an UV divergence. Again,
we choose dimensional regularization to give a meaning to the loop integral.
We work in $d=2\omega$ dimensions, with one temporal and $2\omega-1$
spatial dimensions. 

As in the case of the FY gauge, there is no infrared problem for this term,
and we can put $p=0$ in the photon propagator of the integral (\ref{binan}).
We have

\begin{equation}
e^3\,\int\,\frac{d^{2\omega}k}{(2\pi\mu)^{2\omega}}\,
\frac{\gamma_\nu\,\hat{k}\,
\gamma_\rho\,\hat{k}\,\gamma_\mu\,\,G^{\mu\nu}(k)}
{\left[(-P/2+k)^2-m^2\right]\,\left[(P/2+k)^2-m^2\right]}\, .
\end{equation}

The formulae for integrals of non--covariant functions in
this dimensional regularization prescription can be found in \cite{adkinscoul}.  
After standard computations and taking into account the fact that the terms 
with the Lorentz index $\rho=0$ do not contribute to the amplitude, we obtain:

\begin{equation}
-\frac{i\,e^3}{(4\pi)^2}\,\left(\,\frac{4}{3}\,D\:+\:\frac{20}{9}\,\right)\,
\gamma_\rho 
\label{coultem}
\end{equation}

\noindent
for the case of the temporal propagator, $G_{00}$, and

\begin{equation}
\frac{i\,e^3}{(4\pi)^2}\,\left(\,\frac{1}{3}\,D\:+\:\frac{20}{9}\,\right)\,
\gamma_\rho \label{spacoul}
\end{equation}

\noindent
for the contribution of the spatial components $G_{ij}$ of the
photon propagator. The total result is therefore

\begin{equation}
\frac{\alpha}{4\pi}\,D\,(-i\,e\,\gamma_\rho)\, .
\end{equation}

The contribution of the vertex counterterm is given by 
$\delta Z_1\,(-i\,e\,\gamma_\rho)$ . Since the vertex counterterm to order
$\alpha$ in the Coulomb gauge is \cite{adkinscoul}

\begin{equation}
\delta Z_1\;=\;-\,\frac{\alpha}{4\pi}\,D\, ,
\end{equation}

\noindent
we see that the counterterm exactly cancels the contribution of the second
term in (\ref{totalcoul}), as we wanted to show, and then, the amplitude is 
given by (\ref{dres}) also in the Coulomb gauge.

\subsection{Conclusion.}

As claimed at the beginning of this section, we must consider 
only the subtracted amplitude,

\begin{equation}
M_{\pss AB'}\;=\;M_{\pss AB}\:-\:\frac{\alpha}{\pi} M_{\pss A}\;=\;
-2\,\left(\frac{\alpha}{\pi}\right)^2\,M_{\pss A}\, .
\end{equation}

\noindent

Hence the $\textstyle\protect\rm O_{ps}$ decay rate receives a contribution 
given by the integral of

\begin{equation}
\frac{1}{3}\,\sum_m\,\frac{1}{3!}\sum_\lambda\,2\,{\textstyle Re}\,
\left(M_{\pss AB'}^{\pss (m\,\lambda)}\,^*\,M_0^{(\pss m\,\lambda)}\,\right)\, 
\end{equation}

\noindent
over the phase space of the three final photons. 
Since $M_{\pss AB'}$ is proportional to the lowest order annihilation
amplitude, the contribution to the width is proportional to the lowest order
annihilation width, whose value can be found in \cite{adkins1}:
\, $\Gamma_{\pss A}=-0.81405\, (\alpha/\pi)\, \Gamma_0$ . \\
Therefore, the renormalized vertex correction to the annihilation amplitude 
turns out to be:

\begin{equation}
\Gamma_{\pss AB'}\;=\;-2\,\frac{\alpha}{\pi}\,
\Gamma_{\pss A}\;=\;1.6281\,\frac{\alpha^2}{\pi^2}\,\Gamma_0\, .
\label{result1}
\end{equation}

\noindent
Note that, differently from the lowest order annihilation case, it
contributes positively in the direction of reducing the
theoretical--experimental discrepancy. Its numerical value, however, is 
too small to make a significant progress.

\section{\bf  Lowest Order Annihilation Diagram.}

Now we consider the lowest order annihilation matrix element (fig. 1):

\begin{equation}
-\frac{\sqrt{4\pi\alpha}}{4m^2}\,\int\,\frac{d^4p}{(2\pi)^4}\,Tr\,
\left\{\,\Psi^{\pss (m)}(p)\:
\gamma_{\rho}\,\right\}\:G^{{\pss (\lambda)}\,\rho}\, .
\end{equation}

\noindent
The integral over $p$ is trivially performed. The next stage is to square
the resulting matrix element and make the average over the 
$\textstyle\protect\rm O_{ps}$ polarization states \cite{adkins1}.
After simple calculations, similar to those performed in the paper of one of us
\cite{baier}, ---where the production of tree gluonic jets  in
electron--positron  
colliding beams was considered---, we obtain the following contribution to 
the width\footnote{Remember that the vector $G_\rho^{\pss (\lambda)}$ 
is space-like, as explained before. Therefore, $\Gamma_{\pss A^2}$ is positive,
in spite of what at first sight might seem due to the minus sign in the 
r.h.s. of (\ref{ga2}).}:

\begin{equation}
\Gamma_{\pss A^2}\;=\;- m\,\frac{\alpha^4}{2^{10} \ 9 \ \pi^3}\,
\int\, d^3\nu\:\delta\left(\Sigma\nu_i-2\right)\,\Sigma_{\lambda}\,
G_{\rho}^{\pss (\lambda)}\, G^{{\pss (\lambda)}\,\rho}\, . \label{ga2}
\end{equation}

\noindent
To arrive to (\ref{ga2}), we have expressed the phase space volume of the
three final photons as:

\begin{equation}
\int\,\frac{d^3k_2\,d^3k_3\,d^3k_4}{\omega_2\,\omega_3\,\omega_4}\:
\delta^4(P-k_2-k_3-k_4)\;=\;
8\pi^2m^2\: \int\,d\nu_2\,d\nu_3\,d\nu_4\:\delta\,(2-\nu_2-\nu_3-\nu_4)\, ,
\end{equation}

\noindent
where $\nu_i=\frac{\omega_i}{m} =\frac{|\vec{k}_i|}{m}$ .

At this point we use the results of papers \cite{baier, constan}, namely:

\begin{equation}
-\,\Sigma_{\lambda}\, G_{\rho}^{\pss (\lambda)} \
G^{{\pss (\lambda)}\,\rho}\; = \; 
2^6\,\alpha^4\,\left[\,R(234)+R(324)+R(423)\,\right]\ ;
\end{equation}

\begin{eqnarray}
& & R(234) \; = \; R(243)\;= \nonumber \\
& &\frac{1}{3}\:\left|\,{\cal E}^{(2)}_{-++}(234)\,\right|^2 \;
+ \; \left| \,{\cal E}^{(2)}_{+++}(234)\,\right|^2 \;+\;
\frac{\nu_2}{\nu_3\nu_4 a_2}\:
\left| \,{\cal E}^{(1)}_{-++}(324)\, \right|^2 \; + \nonumber \\ 
& &\frac{1}{\nu_2^2}\: \left|\,{\cal E}^{(1)}_{+++}(234)\:
+ \: {\cal E}^{(1)}_{+++}(243)\,\right|^2
+ \nonumber \\ 
& & \frac{a_3a_4}{\nu_2^2a_2}\: 
\left|\,\frac{1}{a_3}{\cal E}^{(1)}_{+++}(234) \:  -
\: \frac{1}{a_4} {\cal E}^{(1)}_{+++}(243) \, \right|^2\, ,
\end{eqnarray}

\noindent
where $a_i=1-\nu_i$. The rather cumbersome functions $\cal{E}$, whose
arguments are the $\nu_i$, were 
calculated in the paper of Costantini et al. \cite{constan}. Their expressions
are explicitly written 
in appendix B. After numerical integration over the phase space,
we find:

\begin{equation}
\Gamma_{\pss A^2}\;=\;b\,\left(\frac{\alpha}{\pi}\right)^2\,\Gamma_0 \ ,
\end{equation}

\begin{equation}
b\;=\;\frac{\int d^3\nu\,\delta(2-\Sigma_{i}\nu_i)\,
\left[\,R(234)+R(324)+R(423)\,\right]}{32(\pi^2-9)}\;=\;0.17021(10) \ .
\label{result2}
\end{equation}

We would like to stress that the contribution from muon (and hadrons) as 
fermions in the loop is negligible (of order $(m_e/m_\mu)^4$)
\cite{pascual}.

In conclusion, the total correction find here, adding (\ref{result2}) to  
(\ref{result1}), is $1.7983\,(\alpha/\pi)^2\,\Gamma_0$.
Manifestly, it is still far from solving the discrepancy
between the modern theoretical and experimental results. If the $O_{ps}$
problem is to be solved by
 this kind of perturbation theory, larger
contributions to the width must be searched another class of diagrams.

\section{\bf Coulombic Exchange in the Light-Light Scattering Block.}

In this section we shall obtain a known logarithmically enhanced
contribution arising from radiative corrections to the light--light
scattering block. The whole set of these radiative corrections has
been already computed by Adkins and Lymberopoulos \cite{adkins4}. 
We present here a simple computation of the term of order $\alpha^2\ln\alpha$, 
which arises from the diagram displayed in fig. 2--B. 

The contribution to the amplitude of the radiative corrections to the
light--light scattering block can be represented by:

\begin{equation}
M^{\pss (\lambda)}_{\pss AR}\;=\;-i\,e\,\int\,\frac{d^4p}{(2\pi)^4}\,Tr\,
\left\{\,\Psi^{\pss (m)}(p)\,\gamma_\rho\,
\right\}\,\frac{-i}{4m^2}\,G^{{\pss (\lambda)}\,\rho}_{\pss (R)}(k_2,k_3,k_4)\ ,
\end{equation}

\noindent
where the subscript $(R)$ in $G^{{\pss (\lambda)}\,\rho}_{\pss (R)}$ means 
the order $\alpha$ radiative corrections.

We are interested in the contributions to 
$G^{{\pss (\lambda)}\,\rho}_{\pss (R)}$ of the loops
of the form represented by fig. 2--B.\footnote{Strictly speaking, Fig. 2--B
represents only a typical loop. There are five more graphs of the same kind 
corresponding to  permutations of the three final photon states.}
It is given by

\begin{displaymath}
-\,\frac{i\,e^3}{(4\pi)^2}\,\int\frac{d^4l}{(2\pi)^4} 
\int\frac{d^4k}{i\pi^2}\:
\frac{\Delta_{\mu\nu}\,(k-l)}
{(l-k)^2}\;\times
\end{displaymath}
\begin{equation}
Tr \left\{\,
\frac{1}{-\frac{1}{2}\hat{P}+\hat{l}-m}\,\gamma_\nu\,
\frac{1}{-\frac{1}{2}\hat{P}+\hat{k}-m}\,
{\cal O}^{\pss (\lambda)}_{\pss 234}\,\frac{1}{\frac{1}{2}\hat{P}+\hat{k}-m}\,
\gamma_\mu\,
\frac{1}{\frac{1}{2}\hat{P}+\hat{l}-m}\,\gamma_\rho\,\right\}\, .
\label{lungo}
\end{equation}

\noindent
In the previous formula $\Delta_{\mu\nu}(k-l)$ is chosen in the FY gauge 
(\ref{fygauge}) and ${\cal O}^{\pss (\lambda)}_{\pss 234}$ describes the 
annihilation of a pair $e^+\,e^-$ at rest to three photons, namely:

\begin{equation}
{\cal O}^{\pss (\lambda)}_{\pss 234} = -i\,e^3\,\gamma\cdot\epsilon_{\lambda_4}
\, \frac{1}{-\frac{1}{2} 
\hat{P}+\hat{k}+\hat{k_4}-m}\,\gamma\cdot\epsilon_{\lambda_3}\,
\frac{1}{\frac{1}{2} \hat{P}+\hat{k}-\hat{k_2}-m}\,\gamma \cdot 
\epsilon_{\lambda_2} \ .
\label{O234}
\end{equation}

It is worthwhile to underline that there are two regions in the loop
momenta space from which the integral (\ref{lungo}) receives the main 
contributions: one, where the fermion momenta are far off mass--shell, which 
correspond to $l,k\sim m$, and another one with fermion momenta almost  
on mass--shell, $l,k\sim \alpha\, m$. It is this last region which originates
the logarithmic term.
In the analysis of this region we can neglect $k$ and $l$ in the numerators of 
fermion propagators (after rationalizing them) and in 
${\cal O}^{\pss (\lambda)}_{\pss 234}$.
Then, using $\gamma$ matrices algebra, we can rewrite the trace in  
(\ref{lungo}) in the following way :

%\begin{eqnarray}
%Tr\,\left[\,{\cal O}_{234}\,
%\left(\frac{1}{2}\hat{P}+m\right)\, (-ie\gamma_\mu)\, 
%\left(\frac{1}{2}\hat{P}+m\right)\, (-ie\gamma_\rho)\, 
%\left(-\frac{1}{2}\hat{P}+m\right)\,\right. & & \nonumber \\
%\left. (-ie\gamma_\nu)\, \left(-\frac{1}{2}\hat{P}+m\right)\,
%\right]\;=\;-\,e^3\,4\,m^2\,\Delta_{00}(k-l)\,\,
%tr\left[{\cal O}_{234}\,\gamma_\rho
%\right]\,\left(\,1\;+\;O(\alpha^2)\,\right) &  & \ 
%\end{eqnarray}

\begin{equation}
-\,P_\mu P_\nu\,m^2\,Tr\,\left\{\, {\cal O}^{\pss (\lambda)}_{\pss 234}\,
(1+\gamma_0)\,\gamma_\rho\,(1-\gamma_0)\,\right\}\;+\;O(\alpha^2)\, , 
\end{equation}

\noindent
where we have used the fact that $P^2 - 4m^2=O(\alpha^2)$.
The integration in $k$ is now identical to that of (\ref{adkinsint}).
We use the result \cite{adkins1}:

\begin{equation}
\int\,\frac{d^4k}{i\pi^2}\,\frac{-m^2\Delta_{00}(k-l)}
{(k-l)^2\,\left[\,(-\frac{1}{2}P+k)^2 - m^2\,\right]\left[\,(\frac{1}{2}P
+ k)^2 - m^2\,\right]} \approx\frac{\pi\,m}{|\vec{l}|}\,\arctan\frac{|\vec{l}|}
{\gamma}\ - \ 3 .
\end{equation}

The integral in $l$ can be performed in two steps. First, we integrate $l_0$
using the method of residues:

\begin{equation}
\int\,\frac{dl_0}{2\pi}\,\frac{1}{\left[(-\frac{1}{2}P+l)^2-m^2+i0\right]
\left[(\frac{1}{2}P+l)^2-m^2+i0\right]}\;\approx\;
\frac{i}{4 m}\:\frac{1}{|\vec{l}|^2+\gamma^2}\, .
\label{i0eq}
\end{equation}

\noindent
The remaining integral over the spatial 3--momentum $\vec{l}$ is
logarithmically divergent. However, we do not worry here for the 
ultraviolet sector, which is outside the integration region we are considering
and gives no logarithmic contribution in $\alpha$.
 The considerations made at the beginning of this
section permit us to replace the upper limit by $m$ in order to extract
the logarithmic term which arises from the infrared behaviour:

\begin{equation}
\int\,\frac{d^3 l}{(2\pi)^3}\,\frac{i}{4m}\frac{1}{|\vec{l}|^2+\gamma^2}
\,\frac{m\,\pi}{|\vec{l}|}\,\arctan\frac{|\vec{l}|}{\gamma}
\;\approx\;\frac{i}{16}\,\ln\frac{1}{\alpha}\, ,
\end{equation}

\noindent
where we have approximated 
$\arctan \frac{|\vec{l}|}{\gamma}$ \ to \ $\frac{\pi}{2}$. 

Collecting all the factors, we have as a contribution to 
$G_{{\pss (R)}\,\rho}^{\pss (\lambda)}$ :

\begin{equation}
\frac{e^3}{(4\pi)^2}\,\frac{m^2}{4}\,\ln\frac{1}{\alpha}\,\sum_{\sigma\in S_3}
\,Tr\,\left\{\,{\cal O}^{\pss (\lambda)}_{\pss \sigma (234)}\,(1+\gamma_0)
\,\gamma_\rho\,(1-\gamma_0)\,\right\}\, ,
\end{equation}

\noindent
where $\sigma$ is a permutation of $(234)$.
The amplitude receives a contribution:

\begin{eqnarray}
\Delta M_{\pss AR}^{\pss (m\,\lambda)} &=&\alpha^2\,\log\alpha\,\frac{1}{16}\,
\int\frac{d^4p}{(2\pi)^4}\,Tr\,\left\{\,
\Psi^{\pss (m)}(p)\,\gamma_\rho\,\right\}\, \times \nonumber \\
& &\sum_{\sigma\in S_3}\,
Tr\,\left\{\,{\cal O}^{\pss (\lambda)}_{\pss \sigma (234)}\,(1+\gamma_0)
\,\gamma_\rho\,(1-\gamma_0)\,\right\}\, ,
\end{eqnarray}

\noindent
and the corresponding correction to the width is the integral over the three 
final photon phase space of

\begin{equation}
\frac{1}{3}\sum_m\,\frac{1}{3!}\sum_{\lambda}\,2\,{\textstyle Re}\,
\left(\,\Delta M_{\pss AR}^{\pss (m\, \lambda)}\,^*
\,M^{\pss (m\,\lambda)}_0\,\right) .
\end{equation}

\noindent
The lowest order amplitude, $M_0$, can be written in our notation as

\begin{equation}
M^{\pss (m\,\lambda)}_0\;=\;\sum_{\sigma\in S_3}\,\int\,
\frac{d^4p}{(2\pi)^4}\,Tr\,\left\{\,
\Psi^{\pss (m)}(p)\,{\cal O}^{\pss (\lambda)}_{\pss\sigma (234)}\,\right\}\, .
\end{equation}

Using techniques similar to those described in \cite{adkins1},
it is possible to rewrite the product of the traces in such a way that 
the leading width
$\Gamma_0$ appears explicitly. We find 

\begin{equation}
\Delta\,\Gamma_{\pss AR}\;=\;-\,\alpha^2\,\ln\frac{1}{\alpha}\,\Gamma_0 \ .
\end{equation}

\noindent
This result is in agreement with previous calculations 
\cite{lepage, logrus, adkins4}.

\vspace*{1.5cm}
{\sc Acknowledgments}
%\vskip 0.5 truecm
\bigskip

One us (V. A.) would like to thank J\"urg Gasser for useful suggestions.
Another author (E.A.K.) is grateful to the Institute of Theoretical Physics of 
Bern 
Universit\"at for the warm hospitality while doing most of this work.
He would also like to thank I.N. Meshkov for useful discussions. 
V. L. acknowledges  the Ministerio de Educaci\'on y
Cultura (Spain) for finantial support under grant PF--95--73193582.

\appendix
\section{\bf Appendix A.}

In this appendix we discuss the contribution to $M_{\pss AB}$ of the term
$tr_{\mu\nu\rho}(k)-tr_{\mu\nu\rho}(0)$ and of the vertex counterterm, using as
a regulator a sharp cutoff $\Lambda$ in momentum space. Of course the physical
prediction coincides with that obtained in section 2 in dimensional 
regularization.
 
As discussed in section 2, the part of the integral (\ref{binan}) given by
$tr_{\mu\nu\rho}(k)-tr_{\mu\nu\rho}(0)$ has no
infrared singularity and we can put the internal
positronium momentum $p$ equal to zero in the loop integral, 
making only an error of order $\alpha^2$. Hence the integral in $p$ 
can be trivially performed; it remains the loop integral.
Introducing Feynman parameters, making simple integrations 
in\footnote{In this case $\Delta_{\mu\nu}(k)$ is given by (\ref{fygauge}).}

\begin{equation}
-i\,e\frac{\alpha}{4\pi}\int\,\frac{d^4k}{i\pi^2}\,\frac{\gamma_\mu\,\hat{k}\,
\gamma_\rho\,
\hat{k}\,\gamma_\nu\,\Delta_{\mu\nu}(k)}
{k^2\,\left[\,(k+\frac{1}{2}P)^2-m^2\,\right]\left[\,(k-\frac{1}{2}P)^2
-m^2\right]}\, 
\end{equation}

\noindent
and ignoring a term proportional to $P_\rho$, which does not contribute to
the amplitude $M_{AB}$ due to the gauge invariance of the light--light 
scattering tensor, we get  

\begin{equation}
- i\,e\,\frac{\alpha}{4\pi}\,\left(\,3\ln\frac{\Lambda^2}{m^2}\:+\:
\frac{5}{2}\,\right)\,\gamma_\rho\, .
\end{equation}

\noindent
Therefore the contribution of this term to the amplitude is :

\begin{equation}
\frac{\alpha}{4 \pi}\,\left(\,3\ln\frac{\Lambda^2}{m^2}\:+\:
\frac{5}{2}\,\right)\, \left(\frac{\alpha}{\pi} M_{A}\right)\, .
\label{bo}
\end{equation}

The computation of the self--energy ---to get the vertex counterterm--- is
more subtle.  The mass operator to lowest order is: 

\begin{equation}
\Sigma(l)\;=\;-\,i\,\frac{\alpha}{4\pi}\,\int\frac{d^4k}{i\pi^2}\,
\frac{\gamma_{\nu}\,(\hat{l}-\hat{k}+m)\,\gamma_{\mu}\,\Delta_{\mu\nu}(k)}
{(k^2-\lambda^2)\left[(l-k)^2-m^2\right]} \ ,
\end{equation}

\noindent
where $\lambda$ is the fictitious photon mass.
Note that in the FY gauge the numerator does not contain any term which
produces linear divergences, therefore the Feynman trick to join the
denominators and the subsequent shift of the loop momentum can be used,
giving the result:

\begin{eqnarray}
\Sigma(p) &=& \frac{-\,i\,\alpha}{2\pi}\:\int_0^1\,dx\,
\left[\left( 3m-p(1+x) \right) \left( \ln\frac{\Lambda^2}{a^2}-1 \right) \,
-\right. \nonumber \\
&-& \left.\frac{p}{2a^2}\left( (1-x)a^2+4x^2(1-x)p^2 \right) \right]\, ,
\end{eqnarray}

\noindent with

\begin{equation}
a^2\,=\,m^2 \, (x^2+\nu(1-x)+x(1-x)\rho)\, ,\;\;\;\;\;\;\;
\nu=\frac{\lambda^2}{m^2}\, ,\;\;\;\;\;\;\;\rho=1-\frac{p^2}{m^2}\, .
\end{equation}

\noindent
The electron wave function renormalization constant, in the on--mass--shell
renormalization scheme, is defined as  

\begin{equation}
Z=1+\,\left. \frac{d\Sigma(p)}{dp}\,\right|_{\hat{p}=m,\,p^2=m^2} \ .
\end{equation}

\noindent
Simple algebra leads to the following expression for $Z_{FY}^{\pss \Lambda}$: 

\begin{equation}
1+\left. \frac{\alpha}{2\pi}\int_0^1 dx \left[-(1+x)\left(\ln\frac{\Lambda^2}
{m^2}-2\ln x-1\right)
-\frac{9}{2}(1-x)+2J(x,\nu,\rho)\right]\right|_{\nu,\rho \to 0} ,
\end{equation}

\noindent
where

\begin{equation}
J(x,\nu,\rho)=\frac{x\,(1-x)^2\,(2-x)\,(\nu+\rho x)}
{\left[x^2+\nu(1-x)+x(1-x)\rho\right]^2}\ .
\end{equation}

\noindent
The quantity $\int_0^1Jdx$ depends on the way in which $\nu$ and $\rho$ 
tend to zero:

\begin{eqnarray}
\int_0^1Jdx &=& \frac{1}{2}\;\;\;\;\;\;\;\;\;\rho<<\nu \to 0 \ ,\nonumber \\
\int_0^1Jdx &=& 2\;\;\;\;\;\;\;\;\;\nu<<\rho \to 0 \ .
\end{eqnarray}

\noindent
Y. Tomozawa \cite{tomo} showed that the right result appears in the limit 
$\nu<<\rho \to 0$.
The resulting expression is then:
 
\begin{equation}
Z^{\pss \Lambda}_{FY}\;=\;1\:-\:\frac{\alpha}{4\pi}\,
\left(3\,\ln\frac{\Lambda^2}{m^2}-\frac{3}{2}\right)\ .
\end{equation}

\noindent
This implies that the contribution of the vertex counterterm insertion is:

\begin{equation}
-\frac{\alpha}{4\pi}\,\left(\,3\,\ln\frac{\Lambda^2}{m^2}
-\frac{3}{2}\,\right)\, \left(\frac{\alpha}{\pi} M_A \right)\, .
\label{boself}
\end{equation}

To have the total expression of $M_{AB}$ we only need now the contribution of 
the term $tr_{\mu \nu \rho} (0)$. It does not contain any divergence, hence it 
can be recovered directly by eq. (\ref{conbin}) . We have:

\begin{equation}
\left(1 \, - \, 3 \, \frac{\alpha}{\pi} \right) \ \left(\frac{\alpha}{\pi} \, 
M_{A} \right)\ . 
\label{concut}
\end{equation}
 
Collecting the results of (\ref{concut}), (\ref{bo}) and (\ref{boself}),
we get the same as in dimensional regularization (\ref{dres}).

\section{\bf Appendix B.}

We write here the explicit expressions for the quantity $R(234)$ entering
the tensor $G$. Accommodation from the results of paper \cite{constan} 
to the annihilation channel was done in the paper \cite{baier}.

For the case $\epsilon=m$ (notation of \cite{baier}) we obtain:

\begin{eqnarray}
& & {\cal E}^{(1)}_{+++}(234)\; = \nonumber \\
& & \frac{2a_3a_4}{\nu_3}\:+\:
\frac{2a_3}{\nu_3}\,
\left(\frac{2a_3a_4}{a_2}\: +\:\frac{2a_4}{\nu_3} \:-\: a_3\right)\,
\left(B(a_3)-B(1) \right) \; + \nonumber \\
& & 2a_3a_4\,\left(\frac{2}{a_2}\:+\:\frac{1}{\nu_4} \right) \,
\left(B(a_4)-B(1) \right) \; + \nonumber \\
& &\frac{2a_3}{a_2}\, \left(a_3\:-\:a_4\:-\:\frac{2a_3a_4}{a_2}\right)\,
\left(T(a_3)\:+\:T(a_4)\:-\:T(1)\:-\:I(a_3,a_4)\right) \; -
\nonumber \\
& & \frac{a_3}{a_4}\,T(a_2)\;+\;
\frac{a_3}{a_2}\,\left(T(a_2)\:-\:T(a_4)\right) \;-\; T(a_3)\; + \nonumber \\
& &  a_3\,\left(\frac{3}{\nu_3}\:-\:\frac{2a_4}{\nu_3^2}\:-\:\frac{1}{a_2}
\:-\:\frac{1}{a_4}\right)\,\left(T(a_3)\:-\:T(1)\right) \;+ \nonumber \\
& & \frac{\nu_2\,(a_2\,-\,a_4)}{a_2a_4}\,I(a_2,a_3) \;-\; 
\frac{\nu_2 a_3}{a_2 a_4}\,I(a_2,a_4) \;-\; 
\frac{a_3}{\nu_4}\,(T(a_4)\:-\:T(1)) \;+\; \nonumber\\
& & \left(2\:-\:\frac{a_3}{a_4}\:+\:\frac{3a_3}{a_2}\right)\,I(a_3,a_4)\; ; \\
& &   \nonumber \\
& &   \nonumber \\
& & {\cal E}^{(1)}_{-++}(234) \;= \nonumber \\
& & a_3 \,\left(\frac{1}{a_4}\:-\:\frac{1}{a_2}
\right)\,(T(a_2)\:+\:T(a_3)\:+\:T(a_4)\:-\:T(1))\; - \nonumber \\
& &\frac{\nu_2}{a_4}\,I(a_2,a_3)\;+\;\frac{\nu_4}{a_2}\,I(a_3,a_4)\; ; \\
& &   \nonumber \\
& &   \nonumber \\
& & {\cal E}^{(2)}_{+++}(234)\;= \nonumber \\
& &\left(\frac{4a_3}{a_2}\:-\:
\frac{2a_3}{\nu_3}\right)\, \left(B(a_3)\:-\:B(1)\right)\;+\;
\left(\frac{4a_4}{a_2}\:-\:\frac{2a_4}{\nu_4}\right)\,(B(a_4)-B(1))\;-
\nonumber \\
& & \left(\frac{4a_3 a_4}{a_2^2}\:+\:\frac{2\nu_2}{a_2}\right)\,
\left(T(a_3)\:+\:T(a_4)\:-\:T(1)\:-\:I(a_3,a_4)\right)\; - \nonumber \\
& & \left(\frac{1}{a_2}\:+\:\frac{1}{a_3}\:+\:\frac{1}{a_4}\right)\,T(a_2)
\;-\;\left(\frac{a_2}{\nu_3 a_4}\:+\:\frac{3}{a_2}\right)\, T(a_3)\; -
\nonumber \\
& &\left(\frac{a_2}{\nu_4a_3}\:+\:\frac{3}{a_2}\right)\,T(a_4)\;+\;
\left(\frac{\nu_2}{a_3a_4}\:-\:\frac{1}{\nu_3}\:-\:\frac{1}{\nu_4}\:+\:
\frac{3}{a_2}\right)\,T(1)\;+ \nonumber \\
& &\left(\frac{a_4+1}{a_3 a_2}\:+\:\frac{\nu_3}{a_2a_4}\right)\,I(a_2,a_3)
\;+\;\left(\frac{a_3+1}{a_4 a_2}\:+\:\frac{\nu_4}{a_2a_3}\right)\,I(a_2,a_4)
\;+\; \nonumber \\
& & \left(\frac{\nu_2+1}{a_4a_3}\:+\:\frac{5}{a_2}\right)\, I(a_3,a_4)\; ;
\nonumber \\
& &  \nonumber \\
& & \nonumber \\
& & {\cal E}^{(2)}_{-++}(234)\;= \nonumber \\
& & -2\;-\;\left(\frac{1}{a_2}\:+\:\frac{1}{a_3}\:+\:\frac{1}{a_4}\right)
\,(T(a_2)\:+\:T(a_3)\:+\:T(a_4)\:-\:T(1))\;+\; \nonumber \\
& &\left(\frac{1}{a_4}\:+\:\frac{1}{a_2a_3}\right)\,I(a_2,a_3)\;+ 
\left(\frac{1}{a_3}\:+\:\frac{1}{a_2a_4}\right)\,I(a_2,a_4)\;+\;
\nonumber \\
& &\left(\frac{1}{a_2}\:+\:\frac{1}{a_4a_3}\right)\, I(a_3,a_4)\; .
\end{eqnarray}

\noindent
In the previous formulas we used the notations: 

\begin{eqnarray}
a_i &=& 1-\nu_i\: , \: \: \ \nu_2+\nu_3+\nu_4=2\ ,\\
B(z) & = & -1 + \sqrt{\frac{1}{z} - 1}\, \arcsin{\sqrt{z}}\, , \\
T(z) & = & - (\arcsin{\sqrt{z}})^2\, , \\
I(a_3,a_2) & = & F(a_3,\gamma) +  F(a_2,\gamma) - F(1,\gamma)\, , \\
F(a,\gamma) & = & \int_0^1\,\frac{dx}{\gamma^2 - x^2}\,\ln \left[1 - 
a (1-x^2)\right]\, , 
\end{eqnarray}

\noindent with $\gamma = \sqrt{1 + \frac{a_4}{a_3 a_2}}\,$.

\newpage

\begin{figure}[p]
\epsfxsize=10cm
\begin{picture}(50,15) \end{picture}
\epsffile{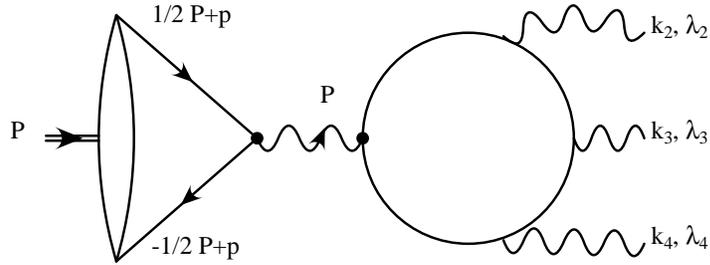}
%\centerline{\epsfbox{fig1.ps}}
\caption{The annihilation graph.}
\end{figure}

\begin{figure}[p]
\epsfxsize=14cm
\epsfysize=5.5cm
\begin{picture}(50,15) \end{picture}
\epsffile{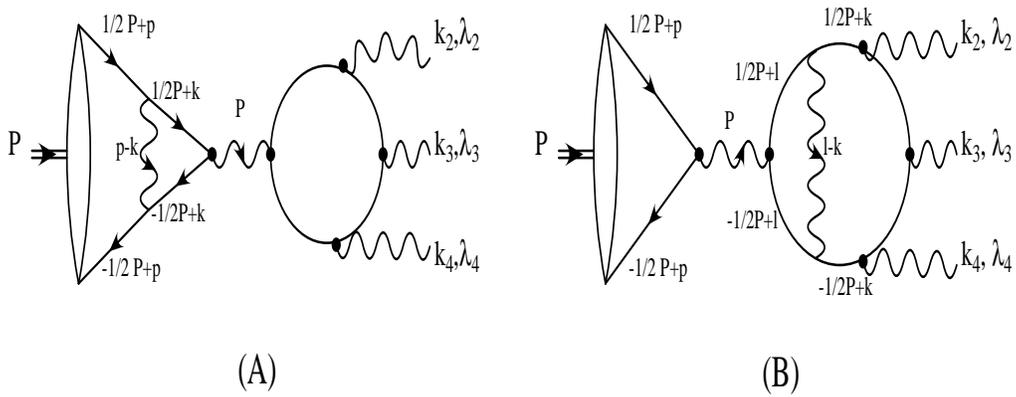}
%\centerline{\epsfbox{nfig2.ps}}
\caption{Two different kinds of corrections to the annihilation graph.
(A) is the vertex correction, (B) represents 
the insertion of a photon into the light--light scattering block, which 
generates a logarithmically enhanced contribution. The direction of the 
fermions in the final loops is clockwise for all the two diagrams.}
\end{figure}

\end{document}